# Retailer response to wholesale stockouts

George Liberopoulos and Isidoros Tsikis

Department of Mechanical Engineering, University of Thessaly, Volos, Greece

November 2012

**Abstract**

**Purpose** – The purpose of this paper is to identify the immediate and future retailer response to wholesale stockouts.

**Design/methodology/approach** – We perform a statistical analysis of historical customer order and delivery data of a local tool wholesaler and distributor, whose customers are retailers, over a period of four years. We investigate the effect of customer service on the order fill rate and the rate of future demand, where the customer service is defined in terms of timely delivery and the fill rate is defined as the fraction of the order that is eventually materialized, i.e., is not cancelled following a stockout.

**Findings** – We find that for customers who order frequently, stockouts have an adverse effect on the fill rate of their orders and on the frequency but not the value of their future demand; however, this latter effect seems to be more short-term than long-term.

**Originality/value** – Practically all studies on the effects of stockouts measure immediate reported/intended consumer purchase incidence and choice decision behavior in response to stockouts in retail environments, mostly based on surveys. This study looks at how stockouts affect future demand in a wholesale environment, based on historical behavioral data analysis.

## Introduction

A stockout occurs whenever an item is demanded from a supplier but can not be delivered because it is temporarily not in stock. In the short run, stockouts may incur backorder and/or lost sales costs. Backorder costs typically include extra costs for administration, price discounts or contractual penalties for late deliveries, expediting material handling and transportation, the potential interest on the profit tied up in the backorder, etc. Lost sales costs include the potential profit loss of the sale if all or part of the sale is lost, contractual penalties for failure to deliver, etc. Besides backorder and lost sales costs, which can be directly measured, a stockout may also incur a less tangible cost in the long run. This cost is related to the loss of customer goodwill. Intuition suggests that a customer that experiences a stockout from a supplier may think twice before placing another order in the future to the same supplier or, even worse, may inform other customers about the disservice he received and influence them into defecting in the future too. In other words, the service level provided by a supplier may influence his future demand and therefore sales. In the short run, sales may fall short of demand when customers experience stockouts and choose not to backorder. In the long run, demand itself may decline as customers who experience excessive stockouts shift temporarily or even permanently to more reliable sources. In general, stockout costs vary at different points in the supply chain (manufacturing, wholesale/distribution, and retail), as the characteristics of customer service and the supplier-customer relationship change along these points.

The quantification of stockout costs has long been a difficult and unsatisfactorily resolved issue in the literature. Nonetheless, the effects of stockouts on customer behavior have been studied quite extensively mostly within the logistics research community and to a lesser extent within the inventory research community. Much of the work reported in the logistics research literature is based on interviews, surveys, and laboratory experiments, mostly on the short-term effects of stockouts, while the work in the inventory management literature focuses on the development and analysis of mathematical inventory models that assume a certain functional dependence of the demand on customer service. Practically all reported studies concern retail environments.

In this paper, we investigate the effect of stockouts on the present and future sales of a firm by performing a statistical analysis of historical customer order and delivery data of a local tool wholesaler and distributor, over a period of four years, where the customers of the wholesaler are retailers. The method that we use is simple. For the nine most important customers of the wholesaler, that account for over 50% of sales, we examine if there is any significant correlation between customer service and the order fill rate, as well as between customer service and the rate of future demand, where by fill rate we mean the fraction of the order that is eventually materialized, i.e., is not cancelled.

Our initial findings are that stockouts have a significant adverse effect on:
1. the order fill rate, for nearly half of the customers,
2. the frequency of future orders, for almost all customers, and

3. the monetary value of future orders, for more than half of the customers.

These initial findings are obtained after applying repeatedly many single-context hypothesis tests, one for each customer. A well-known problem in statistics is that if one performs many such tests, one is likely to find false positives (erroneous significant results). To tackle this problem, we use Holm's (1979) stepdown method to arrive at more conservative conclusions regarding the existence of significant correlations. We also explore if the effects of stockouts on future sales are short-term or long-term.

After applying Holm's method, our more conservative conclusions are that stockouts have a significant adverse effect on:

1. the order fill rate, for the three most frequent customers, and
2. the frequency but not the monetary value of future orders, for two of the three customers of case 1.

Moreover, the latter effect seems to be more short-term than long-term. The customer whose future sales are not affected by stockouts, even though his fill rate is, is the only retail outlet of the wholesaler.

There are very few studies on the effects of stockouts that rely solely on observed order and delivery data and not on data extracted from interviews/surveys. None of these studies concerns a wholesale situation, so in this respect our work adds a contribution to the related literature.

## Literature review

To the best of our knowledge, practically all of the research on the effects of stockouts reported in the logistics research literature has focused on identifying and explaining consumer reaction to stockouts in retail settings. Such reaction may include item (brand and/or variety) or purchase quantity switching, cancellation or deferral of purchase, store switching, etc. Since the late sixties, numerous studies have reported on *intended* consumer behavior response patterns and/or drivers based on interviews and/or surveys. Few studies have investigated *actual* behavior response to stockouts. Of those, some have been based on successive interviews of customers at different points in time, (e.g., before, immediately after, and at some time after a stockout occasion) to compare intended and actual behavior (e.g., Zinn and Liu (2008) and Dadzie and Winston (2007)), while others have investigated the effects of stockouts based on laboratory experiments (e.g., Charlton and Ehrenberg (1976), Motes and Castleberry (1985), and Fitzsimons (2000)).

Most of these works have focused mainly on the immediate impact of stockouts on purchase incidence and choice decisions but have not looked at the cumulative effects of stockouts over time. There exists a very limited number of studies that examine how stockouts affect future long-term demand of retailers, based on historical behavioral data analysis.

Straughn (1991) is one of the first to use scanner data in a stockout study. She attempts to estimate the effects of stockouts on brand share for candy bars. The short-term effect is negligible. The long-term effect, defined as more than five weeks following the stockout condition, is substantial. The decline in brand share averages 10%.

Campo et al. (2003) explore the impact of retail stockouts on whether, how much, and what to buy, by adjusting traditional purchase incidence, quantity and choice models, so as to account for stockout effects. Their study is based on scanner panel data of a large European supermarket chain. They estimate that stockouts may reduce the probability of purchase incidence, lead to the purchase of smaller quantities, and induce asymmetric choice shifts. One limitation of this study is that stockouts are not recorded but are "detected" from the sales data.

Anderson et al. (2006) conduct a large-scale field test with a national mail-order catalog firm and find that stockouts have an adverse impact on both the likelihood that a customer will place another order and the amount that the customer will spend on future orders (if any). They also find that a stockout on one item of an order increases the probability of customers canceling other items in that order, possibly because of the complimentary nature of many of the products sold by the firm, but also because of the shipping and other fixed costs associated with an order.

Jing and Lewis (2011) analyze data from an online grocer to investigate how stockouts affect customers' purchase behavior in both the short- and long-term, how consumers' reactions to stockouts differ across customer segments and how the effect of stockouts varies across product categories. Among others, they find that in the short-run consumers tend to increase buying in categories in which stockouts occur but that stockouts have a negative effect on long-run retention.

Our work follows the limited stream of research that analyzes historical behavioral data to examine the short- as well as the long-term effects of stockouts. Contrary to most other works in the literature that look at retail stockouts, our empirical study focuses on a wholesaler, who – in this case – compared to his retailers, is characterized by fewer recurrent customers that order larger quantities more frequently.

## Data collection

The firm that provided the customer order and delivery data for our study was established as a local retailer of ironware in 1922. Today, it is a wholesaler and distributor of hand tools, hardware, industrial tools and equipment, electric power tools, accessories for power tools, welding machines and accessories, agricultural implements, and other similar products. The firm sells products of many European, Asian, and American tool manufacturers to retailers. The facilities of the firm include a central warehouse that sells items to retailers and a local retail outlet that sells items directly to consumers. The sales department of the firm is staffed with several well-trained salespersons that travel in company owned cars to support its customers (the retailers).

Customers place their orders usually by toll-free phone or fax and sometimes by email, and ideally expect their orders to be met immediately, i.e., within a day. Each order typically contains several items in different quantities and at different prices and is handled by the salesperson who has been assigned to the customer that placed the order. The items of the order that are in stock are delivered to the customer usually on the next working day. Same-day delivery is possible for orders that are placed before noon. The items that are out of stock are backordered. Some of the backordered items may be delivered at a later date or dates. Usually, an order is partially met in more than one delivery, and part of it may be cancelled.

The firm keeps a paper record of every order that is receives, which includes the items that may eventually be cancelled from the order. For each order, it also keeps the delivery invoices with the delivery dates and the items delivered on those dates. We were given limited access to these records for the nine most important customers of the firm that account for over 50% of sales, for a period of four years that included 1043 working days. One of these customers, customer 7, is a retail outlet of the wholesaler.

It is worth mentioning that these records were crucial for our study and we were fortunate that the firm kept them, as we became aware, from discussions with other firms, that in many cases, actual or intended orders are not filed and so stockouts are not recorded. This is also true in many online stores, particularly in cases where the availability of items is indicated in the online catalogue. In such a case, a customer who might want to order an item but sees that it is not available, may not proceed to order it and so the information of this potential order will be lost.

From the records that we collected, we extracted the order and delivery information for each customer. For each order $i$ of each customer, we had access to and collected the following raw data:

$a_i$ : arrival date of the order;

$d_i$ : monetary value of the order that was initially placed, including the value of the items that were eventually cancelled from the order;

$J_i$ : number of deliveries of the order;

$b_{j,i}$ : delivery date of the $j^{th}$ delivery of the order, $j = 1, \ldots, J_i$;

$q_{j,i}$ : monetary value of the $j^{th}$ delivery of the order, $j = 1, \ldots, J_i$;

$\kappa_{j,i} = b_{j,i} - a_i$ : delay in number of working days between the arrival date of the order and the $j^{th}$ delivery date of the order, $j = 1, \ldots, J_i$.

As was mentioned above, our main goal is to examine if the customer service (measured in terms of availability) that the wholesaler provides to any particular order of any particular customer affects:

1. the fill rate of that order, i.e., the fraction of the order that is eventually materialized, (not cancelled), and
2. the rate of future orders of the same customer.

To this end, we defined a set of variables to be used as measures of customer service level, order fill rate, and rate of future orders. More specifically, for each order $i$, we defined the following variables as measures of customer service level and order fill rate, and computed their values:

$x_i = 1 - \sum_{j=1}^{J_i} q_{j,i} / d_i$ : fraction of the value of the order that was cancelled;

$k_i = \kappa_{J_i,i}$ : maximum delivery delay;

$f_i = \sum_{j=1}^{J_i} \kappa_{j,i} (q_{j,i}/d_i) + \beta k_i x_i$ : weighted sum of delivery delays plus a penalty term for the cancelled part of the order;

$s_i$ = stockout occurrence indicator; $s_i = 0$, if $k_i \leq 1$, and $s_i = 1$, if $k_i > 1$.

In the above expression for $f_i$, the summation term represents the weighted sum of the delivery delays, where the delay of each delivery is weighted by the fraction of the value of the order that was filled in that delivery. This term alone, however, does not account for the cancelled part of the order. To account for that part, we assigned to it an artificial delay whose role is analogous to that of the actual delay of a delivered part of the order; namely, the bigger the delay, the worse the customer service. To make this artificial delay weigh relatively heavily on $f_i$, since after all it is associated to a cancellation, we chose it to be equal to the maximum delivery delay, $k_i$, multiplied by a factor of $\beta$, where $\beta > 1$. Then, we multiplied the artificial delay with the fraction of the order that was cancelled and added the resulting product to the summation term in the expression for $f_i$. The results that we report in this paper are for $\beta = 2$. We should note however that when we tried several other values for $\beta$ between 1 and 2, the conclusions that we reached were qualitatively the same as those for $\beta = 2$.

For each order $i$, we also defined the following variables as measures of the rate of future orders, and computed their values:

$e_i = a_{i+1} - a_i$: number of working days until the arrival of the next order;

$h_i = d_{i+1}$: monetary value of the next order placed.

Finally, for each customer we defined the following variables:

$N$: number of orders (observations) recorded;

$M = \sum_{i=1}^{N} s_i$ : number of stockout occurrences;

$E^s = \dfrac{1}{M} \sum_{i=1}^{N} s_i e_i$ : average number of working days until the arrival of the next order following a stockout;

$E^n = \dfrac{1}{N-M} \sum_{i=1}^{N} (1-s_i) e_i$ : average number of working days until the arrival of the next order following a non stockout;

$D^s = \dfrac{1}{M} \sum_{i=1}^{N} s_i d_i$ : average monetary value of the next order placed following a stockout;

$D^n = \dfrac{1}{N-M} \sum_{i=1}^{N} (1-s_i) d_i$ : average monetary value of the next order placed following a non stockout;

## Basic statistical analysis

From the data that we collected for each customer, we computed the sample mean and coefficient of variation (CV) for all the variables; henceforth we will use "$\mu_y$" to denote the mean of any variable $y$. The results are shown in Table 1. The monetary values that appear in Table 1 are the recorded values multiplied by a constant, which is not disclosed here, for confidentiality reasons. The customers are numbered in descending order of their total sales, which is approximately equal to $N \cdot \mu_h$.

From the sample means of the variables of each individual customer shown in Table 1, we can see that the customers exhibited different ordering behaviors in terms of the average frequency and the monetary value of their orders. As a result, they received different average levels of customer service, to which they responded correspondingly. More specifically, we observe the following behavioral patterns.

Customers with larger $\mu_e$ values, i.e., who on average order less frequently, tend to have larger $\mu_h$ values, i.e., tend to place larger orders. The frequency with which a customer orders generally depends on his fixed ordering costs, which to a large extent include the transportation costs for receiving the items and are related to the customer's distance from the supplier. This frequency, however, may also depend on the sophistication of the customer. Customers that are more sophisticated tend to operate in a more "just-in-time" fashion, ordering smaller quantities more frequently.

Customers with larger $\mu_e$ values, i.e., who on average order less frequently, also tend to have larger $\mu_k$ and $\mu_f$ values, i.e., tend to face larger maximum and average delivery delays. This may be due to the fact that these customers can tolerate longer delays, as they include more items in their orders. Customers who order more frequently, on the other hand, may be more pressed to get the items that they request; therefore, they may be less tolerant to long delays and may not wait too long before they switch to alternative sources for the missing items. Although customers with larger interarrival times $\mu_e$ tend to face larger average delivery delays $\mu_f$, these delays, seen as a fraction of the respective $\mu_e$ values, are actually smaller.

Customers with larger $\mu_h$ values, i.e., who on average place larger orders, tend to have larger $\mu_x$ values, i.e., tend to have larger cancellation percentages. This might be explained by the fact that the items demanded by a customer who places larger orders less frequently are perhaps not that crucial to that customer, because they are based on a longer-term – and therefore relatively inaccurate – forecast of his requirements; hence, such a customer might more easily cancel his order for out-of-stock items, irrespectively of the estimated delivery delay for these items. The items demanded by a customer who places smaller orders more frequently, on the other hand, may be more indispensable to that customer, because they are based on a shorter-term – hence, more accurate – forecast of his requirement; therefore he might be more reluctant to drop them from his order.

From the sample CVs of the variables shown in Table 1, we observe that different variables exhibit different levels of variability. Classifying the variability of a random variable as moderate, if its CV is around 1.0, and low/high, if its CV is significantly lower/higher than 1.0, respectively, we can see from the data that the number of days until the arrival of the next order, $e$, and the monetary value of the next order, $h$, has low to moderate variability for all customers. On the other hand, all the variables that are related to customer service and to the customers' immediate response to that service, i.e., $x$, $k$, and $f$, have moderate to high variability for all customers. The fact that the variables related to customer service exhibit higher variability than the variables related to customer demand may be due to the fact that the firm's supply process, which directly affects customer service, is more variable than the demand process. This phenomenon – of increasing variability as one moves upstream the supply chain – is well-known in supply chain management, where it is often referred to as the "bullwhip effect".

**Effect of customer service on present sales**

In the previous section, we conjectured that customers who order less frequently tend to place larger orders and experience longer delivery delays. At the same time, they tend to respond to stockouts with larger order cancellation percentages. To further explore this behavior, we investigated if there is any significant correlation between customer service and the order fill rate for each customer. More specifically, we examined if the maximum delivery delay, $k$, which is a measure of customer service, is significantly correlated with the fraction of the value of the order that is cancelled, $x$, where $x$ is the complement of the order fill rate.

To investigate if there is any significant correlation between $k$ and $x$, we computed Spearman's correlation coefficient $\rho$ which measures the rank-order association between two variables and works regardless of the distributions of the variables. Table 2 shows $\rho$ with its one-tailed significance level $p$ for variables $k$ and $x$, for each customer. Correlations that are significant at a 0.05 level are marked with one asterisk, while those that are significant at a 0.01 level are marked with two asterisks.

From the results displayed in Table 2, we can see that for four out of nine customers, namely customers 2, 6, 7, and 9, there is a significant ($p < 0.05$) positive correlation between $k$ and $x$.

The existence of these correlations implies that when these customers face larger delivery delays, they respond with larger order cancellation percentages.

For the remaining five customers, namely 1, 3, 4, 5, and 8, Spearman's $\rho$ coefficient is positive (except in the case of customer 8) but not significantly ($p < 0.05$) different from zero. For these customers, therefore, there is no significant evidence that the delivery delays affect the order fill rate.

The above analysis is a typical application of multiple hypothesis testing. Namely, for each customer $i$ we tested the null hypothesis $H_i$: "$k_i$ and $x_i$ are not positively correlated," against the alternative hypothesis $\hat{H}_i$: "$k_i$ and $x_i$ are positively correlated." However, as is often noted in the multiple testing literature (e.g., see Westfall and Young (1993)), performing many hypothesis tests may give rise to the "multiple testing problem," which in our case can be stated as follows: The larger the number of customers we perform the test, the more likely we will find significant evidence that $k$ and $x$ is positively correlated for some of these customers, whereas in fact this significance may be due to chance. To tackle the multiple testing problem, and answer the question, "is the significance of the correlation between $k$ and $x$ real or is it due to chance?" we applied Holm's (1979) stepdown method for controlling the family-wise error rate (FWE).

Holm's method works as follows: Order the $p$-values as $p_{(1)} \leq p_{(2)} \leq \ldots \leq p_{(N)}$, where $N$ is the number of test-cases (in our case, customers) and let $H_{(1)},\ldots, H_{(N)}$ denote the corresponding hypotheses, where in our case $H_{(i)}$: "$k_{(i)}$ and $x_{(i)}$ are not correlated." Apply the following sequentially rejective algorithm. If $p_{(1)} > \alpha/N$, *accept* all hypotheses $H_{(1)},\ldots, H_{(N)}$ and *stop*, where $\alpha$ is the preset FWE significance level; otherwise, *reject* $H_{(1)}$ and *continue*. If continuing, then if $p_{(2)} > \alpha/(N-1)$, accept all hypotheses $H_{(2)},\ldots, H_{(N)}$ and *stop*; otherwise, *reject* $H_{(2)}$ and continue; and so on. In general, at the $n$th step, where $n = 1,\ldots, N$, if $p_{(n)} > \alpha/(N - n + 1)$, *accept* all hypotheses $H_{(n)},\ldots, H_{(N)}$ and *stop*; otherwise, reject $H_{(n)}$ and *continue* to the next step.

Applying Holm's method to the data displayed in Table 2 leads to the following conclusion: For both $\alpha = 0.01$ and $\alpha = 0.05$:

– $k$ and $x$ are positively correlated for customers 2 and 7, and
– $k$ and $x$ are not correlated for the remaining customers.

We should note, however, that while the FWE is strongly protected using Holm's step-down method, it is based on the Bonferroni probability inequality, and hence is conservative, i.e., it is more difficult to lead to a "reject $H_n$" conclusion. Note that if we apply Holm's method for $\alpha =$ 0.06, customer 9 will also join the list of customers for which we can accept the hypothesis that $k$ and $x$ are positively correlated.

To summarize, after applying Holm's method, we can conservatively conclude that $k$ and $x$ are positively correlated for three customers at the 0.06 significance level. Looking at Table 1, these three customers, namely 2, 7, and 9, are the customers with the smallest mean $e$ values, i.e., they are those who on average order more frequently. In addition, they are those that exhibit the smallest difference between mean maximum delivery delay, $\mu_k$, and mean order interarrival time, $\mu_e$. In other words, the maximum delivery delay $k$ is positively correlated with the cancellation percentage $x$ for the most frequent customers. This may be due to the fact that frequent customers are more pressed to receive the out-of-stock items. Although they may be reluctant to cancel these items, because they need them to fill their short-term – hence, relatively accurate – requirements, they may not hesitate to drop them from their order and look for them elsewhere, if the anticipated delivery time is greater that the time of their next order.

**Effect of customer service on future sales**

In the previous section, we concluded that stockouts had a significant adverse effect on the fill rate of customers who order frequently. The next question that we posed is whether stockouts also undermine future sales. To answer this question, we investigated if any of the variables that measure the magnitude of stockouts, which we refer to as *independent* variables, were significanty correlated with the variables that measure the change in the rate of future customer orders, which we refer to as *dependent* variables.

The independent variables that measure the magnitude of a stockout faced by any particular order $i$ of any particular customer are the cancellation percentage $x_i$, the maximum delivery delay $k_i$, and the weighted sum of delivery delays $f_i$. The dependent variables that measure the change in the rate of future customer orders following order $i$ are the elapsed time until the next order $e_i$ and the value of the next order $h_i$. Intuition suggests that a drop in the rate of future customer orders may be affected not only by the most recent stockout experienced by a customer but by previous stockouts as well, although the effect of older stockouts on the drop in future customer demands should be less intense than the effect of more recent stockouts. In order to test the hypothesis that the drop – if any – in the rate of future customer orders due to the loss of customer goodwill is a phenomenon that is cumulative over time but at the same time customers are forgetting or forgiving as time passes, we introduced four new sets of variables, which were defined as the exponentially smoothed versions of the four original independent variables, $x_i$, $k_i$, and $f_i$. In each new variable, the magnitude of the stockout that the customer faced on his $i^{th}$ order is measured by weighing the current value as well as all the previous values of the respective variable with geometrically decreasing weights as we go back in time. More specifically, the exponentially smoothed versions of the independent variables were defined as follows:

$$X_i^{\gamma} = \gamma x_i + (1-\gamma) X_{i-1}^{\gamma},$$

$$K_i^{\gamma} = \gamma k_i + (1-\gamma) K_{i-1}^{\gamma},$$

$$F_i^{\gamma} = \gamma f_i + (1-\gamma) F_{i-1}^{\gamma},$$

where $\gamma$ is the smoothing factor. Note that as $\gamma$ tends to 1, more weight is being placed on the more recent value of the independent variable, whereas as $\gamma$ tends to 0, more weight is being placed on past values of the independent variable. In this study, we considered four values for $\gamma$, namely, 0.2, 0.4, 0.6, 0.8, and 1.

We computed Spearman's correlation coefficient $\rho$ with its one-tailed significance level $p$ for each pair of independent variables, $X^{\gamma}$, $K^{\gamma}$, and $F^{\gamma}$, and dependent variables, $e$ and $h$, for $\gamma = 0.2$, 0.4, 0.6, 0.8, and 1. The results are shown in Tables 3 and 4, where the correlations that are significant at a 0.05 level are marked with one asterisk, while those that are significant at a 0.01 level are marked with two asterisks.

From the results displayed in Table 3, we can see that for eight out of nine customers at least one of the independent variables shows a significant ($p < 0.05$) correlation with dependent variable $e$. The only exception is customer 7, for whom no significant correlation is found. Also, from the results displayed in Table 4, we can see that for five out of nine customers, namely customers 2, 4, 5, 6, and 8, at least one of the independent variables shows a significant ($p < 0.05$) correlation with dependent variable $d$. Moreover, for the majority of the cases that show significant correlation between an independent and a dependent variable, the corresponding correlation coefficient $\rho$ is below 0.4, indicating that this correlation is not very strong.

From the data in Table 3, we can see that for all but one cases which show a significant correlation between an independent variable and variable $e$, this correlation is positive. This is in line with intuition which suggests that the larger the value of the independent variable, the lower the service level, and hence the longer the time until the next order, $e$. The only case where a significant correlation coefficient is negative is the case of the coefficient between $X^{0.2}$ and $e$, for customer 5. In fact, this coefficient has the largest absolute value (0.3830) among all coefficients.

Its negative sign, however, is counter intuitive and raises the suspicion that its apparent significance may be due to chance.

Moreover, from the results shown in Table 4, for the cases which show a significant correlation between an independent variable and variable $h$, this correlation is sometimes negative and sometimes positive. A negative correlation coefficient means that the lower the customer service level in a stockout situation, the smaller the monetary value of the order following the stockout. This type of behavior is also reported in Campo et al. (2003) and Anderson et al. (2006), who find that customers that experience stockouts, spend less money (i.e., place orders of smaller monetary value) following the stockouts, although in both these studies this effect is small. A positive correlation coefficient, on the other hand, means that the lower the service level in a stockout situation, the larger the monetary value of the order following the stockout. This type of behavior is counter-intuitive. One possible explanation for it is that a customer who faces a stockout may delay his order following that stockout, because of the dissatisfaction, but when he returns, he orders a larger amount, because his requirements have increased in the mean time (assuming of course that he has not satisfying all of his requirements elsewhere). Nevertheless, this wavering behavior again raises the suspicion that the respective significance may be due to chance.

To answer the question "are the observed significances real or are they due to chance?" we applied again Holm's stepdown method to the data shown in Tables 3 and 4. The conclusions are:

For $\alpha = 0.01$:
- Independent variables $F^1$ and $F^{0.8}$ are positively correlated with $e$ for customer 2,
- independent variables $K^1$, $F^1$, and $F^{0.8}$ are positively correlated with $e$ for customer 9, and
- none of the remaining independent variables is correlated with $e$.

For $\alpha = 0.05$:
- Independent variables $X^1$, $X^{0.8}$, $X^{0.6}$, $F^1$, $F^{0.8}$, $F^{0.6}$, and $F^{0.4}$ are positively correlated with $e$ for customer 2,
- independent variables $X^1$, $K^1$, $K^{0.8}$, $K^{0.6}$, $F^1$, and $F^{0.8}$ are positively correlated with $e$ for customer 9, and
- none of the remaining independent variables is correlated with $e$.

For both $\alpha = 0.01$ and $\alpha = 0.05$:

– none of the independent variables is correlated with *h* for any customer.

To summarize, after applying Holm's method, we can conservatively conclude that for customers 2 and 9, some of the independent variables measuring the magnitude of a stockout are positively correlated with the time until the next order following the stockout, *e*, at both the 0.01 and 0.05 significance levels. Moreover, for these two customers, wherever there is a significant correlation between an independent variable and the dependent variable *e*, the higher the value of the smoothing factor $\gamma$, the bigger the correlation coefficient. This suggests that wherever *e* is significantly affected by a stockout, it is mostly affected by the most recent stockout than by previous stockouts. In fact, for $\gamma = 0.2$, no independent variable is correlated with *e*, for neither customer. This suggests that the adverse effect of a stockout on future demand is short-term.

We can also conservatively conclude that none of the independent variables measuring the magnitude of a stockout are correlated with the monetary value of the order following the stockout, *h*, at either the 0.01 or the 0.05 significance level, for any customer. This means that, although some customers who experience a stockout may delay their next order following a stockout, none reduces the monetary value of his next order.

In the previous section, we conservatively concluded that for the three most frequent customers, namely customers 2, 7, and 9, stockouts have a significant effect on the order fill rate. In this section, we conservatively concluded that for two of these customers, namely customers 2 and 9, stockouts also have a significant effect on the frequency but not on the monetary value of future orders and that this effect seems to be more short- than long-term. This is not true for customer 7 who is the only retailer that is actually owned by the wholesaler. Therefore, even though a stockout may force this customer to cancel the missing part of an order and look for it elsewhere, it does not affect his future demand and hence loyalty to the wholesaler, neither in the long- nor in the short-term. It is interesting to note from Table 1, that customer 7 has the smallest absolute difference between mean interarrival time, $\mu_e$, and maximum delivery delay, $\mu_k$. This suggests that even when he cancels the missing part of an order, he does so just a little while before he places his next order, i.e., "at the last moment".

## Summary and implications of results

Our empirical findings from this case study and some of their implications for researchers and practitioners can be summarized as follows.

Customers who order less frequently tend to place larger orders and tolerate longer delivery delays. At the same time, they tend to respond to stockouts with larger order cancellation percentages, most likely because the out-of-stock items are not that indispensable to them, since their orders are based on longer-term – hence, less accurate – forecasts. This suggests that inventory control models in which order fill rates are assumed to depend on order frequencies may be a good representation of reality in environments similar to ours.

The variables that are related to customer service exhibit higher variability than the variables related to customer demand. This is probably due to the fact that the firm's supply process, which directly affects customer service, is more variable than the demand process. The elevated variability in customer service is to some extent due to the highly disruptive effect of stockouts. A better design of the stocking and reordering policy used by the company might help reduce some of this variability.

Stockouts do have an adverse effect on present and future sales for customers who order frequently. They have an adverse effect on present sales, most likely because frequent customers are more pressed to receive the out-of-stock items. They seem to be reluctant to cancel these items, because they need them to fill their short-term – hence, relatively accurate – requirements, but at the same time, they do not wait long before they look for them elsewhere.

There are three possible explanations for why stockouts have an adverse effect on future sales for frequent customers.

1. When a frequent customer places an order, he has a more vivid memory of his previous orders and associated stockouts than a less frequent customer, simply because these orders are more recent. Our analysis shows, however, that even for frequent customers this memory seems to be short-term and does not affect long-term sales.

2. When a frequent customer is forced to demand the missing items of an order from an alternative supplier, the instant that he places his demand to that supplier is often very close to the instant that he is about to place his next order to the original wholesaler. In such a case, he may decide to skip placing his next order to the original wholesaler and instead place it to the alternative supplier along with his demand for the missing items, to save fixed order costs. Indeed, the evidence in Table 1 confirms that for frequent customers, the mean value of the maximum delivery delay, $\mu_k$, which should be related to the time that a customer decides to cancel the missing part of his order, is quite close to the mean interarrival time, $\mu_e$.

3. A frequent customer may operate in a more "just-in-time" manner than a less frequent customer, in which case he van be considered to be more flexible and sophisticated. Being more sophisticated, he is more sensitive and reactive to poor customer service on the part of the wholesaler.

There are several issues that we did not take into account in this study. We did not take into account factors for which we had no data, such as the detailed list of items in each order, whether the customers accepted item substitution in case of a stockout, whether the firm offered a price discount for the out-of-stock items, whether the cancelled items of an order were purchased from an alternative wholesaler or were included in a subsequent order, etc. Future research should be directed towards including such details in the analysis. It could also address the issue of availability of items from alternative suppliers which would impact customer response to stockouts.

# Tables

Table 1: Statistics of the order and delivery variables

| | | Customer | | | | | | | | |
|---|---|---|---|---|---|---|---|---|---|---|
| | | 1 | 2 | 3 | 4 | 5 | 6 | 7 | 8 | 9 |
| | $N$ | 53 | 121 | 80 | 41 | 42 | 59 | 147 | 48 | 247 |
| Variable | Statistic | | | | | | | | | |
| $x$ | Mean | 0.0543 | 0.0453 | 0.0310 | 0.0351 | 0.0803 | 0.0473 | 0.0447 | 0.0271 | 0.0180 |
| | CV | 1.5816 | 2.8148 | 3.1807 | 2.5932 | 1.4645 | 1.7937 | 2.2196 | 3.6385 | 3.7844 |
| $k$ | Mean | 13.377 | 4.0331 | 3.1875 | 14.293 | 14.119 | 5.0508 | 4.5714 | 9.2500 | 0.6761 |
| | CV | 1.1937 | 2.0283 | 2.6324 | 1.3330 | 1.0503 | 1.9536 | 5.2273 | 1.4019 | 3.4418 |
| $f$ | Mean | 4.9908 | 2.9352 | 2.9795 | 4.6097 | 6.0301 | 3.0252 | 2.5095 | 4.5409 | 1.7865 |
| | CV | 1.1002 | 2.0763 | 2.2602 | 1.9171 | 0.9478 | 1.1370 | 1.9906 | 1.8950 | 5.9746 |
| $e$ | Mean | 17.943 | 5.9256 | 8.4000 | 18.683 | 18.595 | 15.797 | 4.7415 | 18.917 | 2.8300 |
| | CV | 0.7302 | 0.8228 | 0.9012 | 0.6554 | 0.6010 | 0.6473 | 0.7475 | 0.7244 | 0.7489 |
| $h$ | Mean | 1749.4 | 443.08 | 509.87 | 890.08 | 868.07 | 500.02 | 181.84 | 403.29 | 76.352 |
| | CV | 0.9304 | 1.1447 | 0.8279 | 0.8939 | 0.7317 | 0.6552 | 1.1420 | 0.9094 | 0.7363 |
| | $M$ | 21 | 38 | 43 | 21 | 25 | 35 | 27 | 30 | 25 |
| | $E^s$ | 19.4211 | 7.1860 | 12.667 | 20.3600 | 18.6000 | 19.3793 | 5.9667 | 21.5862 | 3.7037 |
| | $E^n$ | 14.2000 | 5.2308 | 6.8814 | 16.0625 | 18.5714 | 12.3333 | 4.4274 | 14.8421 | 2.7227 |
| | $D^s$ | 1587.41 | 323.42 | 527.22 | 808.969 | 781.785 | 537.550 | 135.61 | 383.000 | 69.800 |
| | $D^n$ | 2159.95 | 509.04 | 503.69 | 1016.82 | 1299.11 | 463.742 | 193.69 | 434.257 | 77.156 |

Table 2: Spearman's $\rho$ correlation coefficient and corresponding one-tailed significance level $p$ regarding the correlation between variables $k$ and $x$

| | Customer | | | | | | | | |
|---|---|---|---|---|---|---|---|---|---|
| | 1 | 2 | 3 | 4 | 5 | 6 | 7 | 8 | 9 |
| $\rho$ | 0.2206 | 0.4915** | 0.1155 | 0.0903 | 0.0867 | 0.2566* | 0.4133** | -0.0769 | 0.1532* |
| $p$ | 0.0562 | 0.0000 | 0.1539 | 0.2873 | 0.2925 | 0.0249 | 0.0000 | 0.3017 | 0.0082 |

Table 3: Spearman's $\rho$ correlation coefficient and corresponding one-tailed significance level $p$ regarding the correlation between each independent variable and variable $e$

| Ind. var. | | Customer 1 | 2 | 3 | 4 | 5 | 6 | 7 | 8 | 9 |
|---|---|---|---|---|---|---|---|---|---|---|
| $X^1$ | $\rho$ | 0.0453 | 0.2658** | 0.0054 | 0.0859 | 0.0083 | -0.1206 | -0.0208 | 0.2768* | 0.1647** |
| | $p$ | 0.3738 | 0.0016 | 0.4810 | 0.2967 | 0.4792 | 0.1814 | 0.4013 | 0.0284 | 0.0048 |
| $X^{0.8}$ | $\rho$ | 0.0234 | 0.2487** | 0.0415 | 0.1907 | -0.0494 | -0.0852 | 0.0079 | 0.1464 | 0.0533 |
| | $p$ | 0.4339 | 0.0030 | 0.3575 | 0.1161 | 0.3780 | 0.2605 | 0.4622 | 0.1603 | 0.2023 |
| $X^{0.6}$ | $\rho$ | 0.1087 | 0.2305** | 0.0422 | 0.2027 | -0.1573 | -0.0554 | 0.0312 | 0.1387 | 0.0257 |
| | $p$ | 0.2193 | 0.0055 | 0.3551 | 0.1019 | 0.1599 | 0.3385 | 0.3536 | 0.1735 | 0.3441 |
| $X^{0.4}$ | $\rho$ | 0.1708 | 0.1998* | 0.0618 | 0.1852 | -0.2397 | -0.0185 | 0.0497 | 0.1180 | -0.0196 |
| | $p$ | 0.1107 | 0.014 | 0.2930 | 0.1231 | 0.0631 | 0.4447 | 0.2752 | 0.2121 | 0.3796 |
| $X^{0.2}$ | $\rho$ | 0.1972 | 0.1836* | 0.0886 | 0.1583 | -0.3830** | 0.0042 | 0.0985 | 0.0558 | -0.1017 |
| | $p$ | 0.0785 | 0.0219 | 0.2173 | 0.1615 | 0.0061 | 0.4875 | 0.1177 | 0.3531 | 0.0555 |
| $K^1$ | $\rho$ | 0.1200 | 0.1595* | 0.1904* | 0.2368 | 0.1937 | 0.3064** | 0.0888 | 0.2844* | 0.2232** |
| | $p$ | 0.1959 | 0.0403 | 0.0454 | 0.0680 | 0.1095 | 0.0091 | 0.1424 | 0.0250 | 0.0002 |
| $K^{0.8}$ | $\rho$ | 0.2198 | 0.1001 | 0.1932* | 0.2638* | 0.2625* | 0.3043** | 0.0840 | 0.1691 | 0.1793** |
| | $p$ | 0.0569 | 0.1374 | 0.0430 | 0.0478 | 0.0465 | 0.0096 | 0.1558 | 0.1253 | 0.0023 |
| $K^{0.6}$ | $\rho$ | 0.2483* | 0.1015 | 0.2010* | 0.2336 | 0.2782* | 0.2947* | 0.0897 | 0.158 | 0.1672** |
| | $p$ | 0.0365 | 0.1341 | 0.0369 | 0.0708 | 0.0372 | 0.0117 | 0.1399 | 0.1417 | 0.0042 |
| $K^{0.4}$ | $\rho$ | 0.2914* | 0.1232 | 0.1597 | 0.1702 | 0.2943* | 0.2815* | 0.1104 | 0.1290 | 0.1353* |
| | $p$ | 0.0171 | 0.0892 | 0.0785 | 0.1436 | 0.0293 | 0.0154 | 0.0915 | 0.1911 | 0.0168 |
| $K^{0.2}$ | $\rho$ | 0.2369* | 0.2044* | 0.1515 | 0.0578 | 0.2762* | 0.2058 | 0.1321 | 0.0859 | 0.1281* |
| | $p$ | 0.0438 | 0.0122 | 0.0899 | 0.3598 | 0.0383 | 0.0589 | 0.0554 | 0.2808 | 0.0222 |
| $F^1$ | $\rho$ | 0.1303 | 0.2606** | 0.1126 | 0.1323 | 0.0959 | 0.1765 | -0.0048 | 0.2982* | 0.2677** |
| | $p$ | 0.1762 | 0.0019 | 0.1599 | 0.2047 | 0.2729 | 0.0906 | 0.4768 | 0.0198 | 0.0000 |
| $F^{0.8}$ | $\rho$ | 0.1143 | 0.2561** | 0.1264 | 0.2153 | 0.0724 | 0.1920 | 0.0298 | 0.2618* | 0.1938** |
| | $p$ | 0.2075 | 0.0023 | 0.1320 | 0.0882 | 0.3244 | 0.0725 | 0.3601 | 0.0362 | 0.0011 |
| $F^{0.6}$ | $\rho$ | 0.1114 | 0.239** | 0.1423 | 0.2758* | 0.0173 | 0.2003 | 0.0333 | 0.2296 | 0.1410* |
| | $p$ | 0.2135 | 0.0041 | 0.1040 | 0.0404 | 0.4568 | 0.0642 | 0.3446 | 0.0583 | 0.0134 |
| $F^{0.4}$ | $\rho$ | 0.1107 | 0.2439** | 0.1422 | 0.3079* | -0.0411 | 0.1442 | 0.0557 | 0.1683 | 0.0678 |
| | $p$ | 0.2150 | 0.0035 | 0.1041 | 0.0251 | 0.3980 | 0.1379 | 0.2515 | 0.1264 | 0.1441 |
| $F^{0.2}$ | $\rho$ | 0.0293 | 0.2139** | 0.1637 | 0.2961* | -0.0315 | 0.0719 | 0.0945 | 0.1181 | -0.0103 |
| | $p$ | 0.4176 | 0.0092 | 0.0734 | 0.0301 | 0.4216 | 0.2942 | 0.1275 | 0.2120 | 0.4359 |

Table 4: Spearman's $\rho$ correlation coefficient and corresponding one-tailed significance level $p$ regarding the correlation between each independent variable and variable $h$

| Ind. var. | | \multicolumn{9}{c}{Customer} | | | | | | | | |
|---|---|---|---|---|---|---|---|---|---|---|
| | | 1 | 2 | 3 | 4 | 5 | 6 | 7 | 8 | 9 |
| $X^1$ | $\rho$ | -0.1469 | -0.1725* | 0.1584 | 0.1368 | -0.2703* | -0.1992 | -0.0414 | 0.3536** | 0.0086 |
| | $p$ | 0.1470 | 0.0292 | 0.0803 | 0.1969 | 0.0417 | 0.0652 | 0.3092 | 0.0068 | 0.4465 |
| $X^{0.8}$ | $\rho$ | -0.1386 | -0.1162 | 0.1748 | 0.3791** | -0.3231* | -0.1694 | -0.0354 | -0.0231 | -0.0249 |
| | $p$ | 0.1611 | 0.1023 | 0.0604 | 0.0073 | 0.0184 | 0.0998 | 0.3353 | 0.4380 | 0.3487 |
| $X^{0.6}$ | $\rho$ | -0.1029 | -0.1017 | 0.1793 | 0.3887** | -0.3688** | -0.1503 | -0.0192 | -0.0354 | -0.0192 |
| | $p$ | 0.2317 | 0.1335 | 0.0557 | 0.006 | 0.0081 | 0.1279 | 0.4089 | 0.4056 | 0.3819 |
| $X^{0.4}$ | $\rho$ | -0.0518 | -0.0800 | 0.1627 | 0.3404* | -0.3049* | -0.0891 | 0.0054 | -0.0207 | -0.0087 |
| | $p$ | 0.3562 | 0.1916 | 0.0747 | 0.0147 | 0.0248 | 0.2510 | 0.4741 | 0.4444 | 0.4456 |
| $X^{0.2}$ | $\rho$ | -0.0877 | -0.0531 | 0.1625 | 0.2767* | -0.2910* | 0.0038 | 0.0349 | -0.0390 | 0.0059 |
| | $p$ | 0.2661 | 0.2816 | 0.0749 | 0.0400 | 0.0308 | 0.4886 | 0.3374 | 0.3963 | 0.4634 |
| $K^1$ | $\rho$ | -0.0666 | -0.2118** | 0.0262 | -0.1012 | -0.1508 | 0.0849 | -0.1345 | -0.1044 | 0.0098 |
| | $p$ | 0.3179 | 0.0099 | 0.4086 | 0.2644 | 0.1703 | 0.2613 | 0.0522 | 0.2401 | 0.4389 |
| $K^{0.8}$ | $\rho$ | 0.0028 | -0.181* | -0.0039 | -0.1078 | -0.1036 | 0.1589 | -0.1186 | -0.0823 | -0.0626 |
| | $p$ | 0.4920 | 0.0235 | 0.4862 | 0.2511 | 0.2568 | 0.1147 | 0.0762 | 0.2891 | 0.1637 |
| $K^{0.6}$ | $\rho$ | 0.0331 | -0.1657* | -0.0234 | -0.1162 | -0.1059 | 0.1963 | -0.1209 | -0.0587 | -0.0681 |
| | $p$ | 0.4069 | 0.0347 | 0.4182 | 0.2347 | 0.2522 | 0.0681 | 0.0723 | 0.3459 | 0.1430 |
| $K^{0.4}$ | $\rho$ | 0.0235 | -0.1610* | -0.0665 | -0.1831 | -0.0978 | 0.2183* | -0.1174 | -0.0322 | -0.0883 |
| | $p$ | 0.4338 | 0.0389 | 0.2790 | 0.1259 | 0.2689 | 0.0484 | 0.0784 | 0.4139 | 0.0833 |
| $K^{0.2}$ | $\rho$ | 0.0090 | -0.1565* | -0.0465 | -0.1887 | -0.1524 | 0.1449 | -0.0935 | -0.0208 | -0.0752 |
| | $p$ | 0.4746 | 0.0432 | 0.3410 | 0.1187 | 0.1676 | 0.1367 | 0.1299 | 0.4441 | 0.1194 |
| $F^1$ | $\rho$ | 0.0346 | -0.2147** | 0.1342 | -0.0262 | -0.0640 | 0.0949 | -0.0937 | -0.1253 | 0.0511 |
| | $p$ | 0.4030 | 0.0090 | 0.1177 | 0.4353 | 0.3437 | 0.2373 | 0.1296 | 0.1981 | 0.2122 |
| $F^{0.8}$ | $\rho$ | 0.0223 | -0.1987* | 0.1561 | 0.1129 | -0.0913 | 0.0749 | -0.0781 | -0.1202 | 0.0568 |
| | $p$ | 0.4372 | 0.0144 | 0.0834 | 0.2411 | 0.2826 | 0.2865 | 0.1734 | 0.2079 | 0.1870 |
| $F^{0.6}$ | $\rho$ | -0.0109 | -0.1827* | 0.1220 | 0.1294 | -0.1314 | 0.0666 | -0.0736 | -0.0485 | 0.0588 |
| | $p$ | 0.4692 | 0.0224 | 0.1404 | 0.2099 | 0.2035 | 0.3082 | 0.1877 | 0.3716 | 0.1786 |
| $F^{0.4}$ | $\rho$ | -0.0252 | -0.1444 | 0.0903 | 0.1564 | -0.1066 | 0.0568 | -0.0513 | -0.0423 | 0.0676 |
| | $p$ | 0.4290 | 0.0571 | 0.2127 | 0.1643 | 0.2509 | 0.3346 | 0.2686 | 0.3876 | 0.1448 |
| $F^{0.2}$ | $\rho$ | -0.0731 | -0.1024 | 0.0924 | 0.1986 | -0.1529 | 0.1084 | -0.0270 | -0.0244 | 0.1023 |
| | $p$ | 0.3016 | 0.1318 | 0.2075 | 0.1066 | 0.1668 | 0.2069 | 0.3728 | 0.4346 | 0.0543 |